# 21$^{st}$ century change in ocean response to climate forcing

Stjepan Marčelja[*]

Department of Applied Mathematics, Research School of Physics and Engineering,
Australian National University, Canberra ACT 2601, Australia

**Abstract**

Modeling globally averaged information on climate forcing from the land surface temperature data, the sea surface temperatures (SST) and the empirically determined relationship between the changes in SST and the turbulent diffusion of heat into the upper ocean demonstrates a consistent link. The modeling is accurate throughout the 20$^{th}$ century despite the different phases of the Interdecadal Pacific Oscillation (IPO) or the strong divergence between land and ocean surface warming. It only fails during the last 15 years when SST drops well below the trend. The finding reinforces the view that slower global warming over the previous 15 years is not a caused by a negative phase of the IPO or by the variations in the upper ocean (top 700 m) warming but results from a change in the ocean behavior leading to increased heat transfer into the deeper ocean.

## 1. Introduction

Changes in global land and ocean surface temperature show remarkable concurrence throughout the 80-year period starting in 1900. However, since about 1980 the land and ocean surface warming rates have dramatically diverged (Fig. 1a). While the 20$^{th}$ century periods of rapid or slow warming can be loosely correlated with the phase of the IPO oscillation (Meehl et al., 2013, 2014), the same is not true for the land-ocean temperature divergence. Different rates of warming, which are a consequence of differences in the heat capacity of land and ocean, emerged prominently at the time of intensifying climate forcing. This changed feature in global climate behavior has received comparatively little attention, even though such long and accurate records should provide insight into the nature of the slowdown in global warming.

[*] Email: stjepan.marcelja@anu.edu.au
Phone: +61 475 89 6595

While the land surface temperature anomaly is almost proportional to the climate forcing, SST anomaly is more involved because of the large heat transfer into the upper ocean and the poorly known heat transfer into the deep ocean. The slowdown in global surface warming is predominantly caused by the weaker increases in SST since about the year 2000 (Fig. 1a). The information contained in the link between the SST and the climate forcing (as reflected in land surface temperatures) is therefore able to provide insight into the causes of the slowdown. In particular, we wish to consider the possibility that the slower warming is a part of a systemic change in the global response to the increasing climate forcing rather then a passing climate fluctuation such as a negative phase of the IPO.

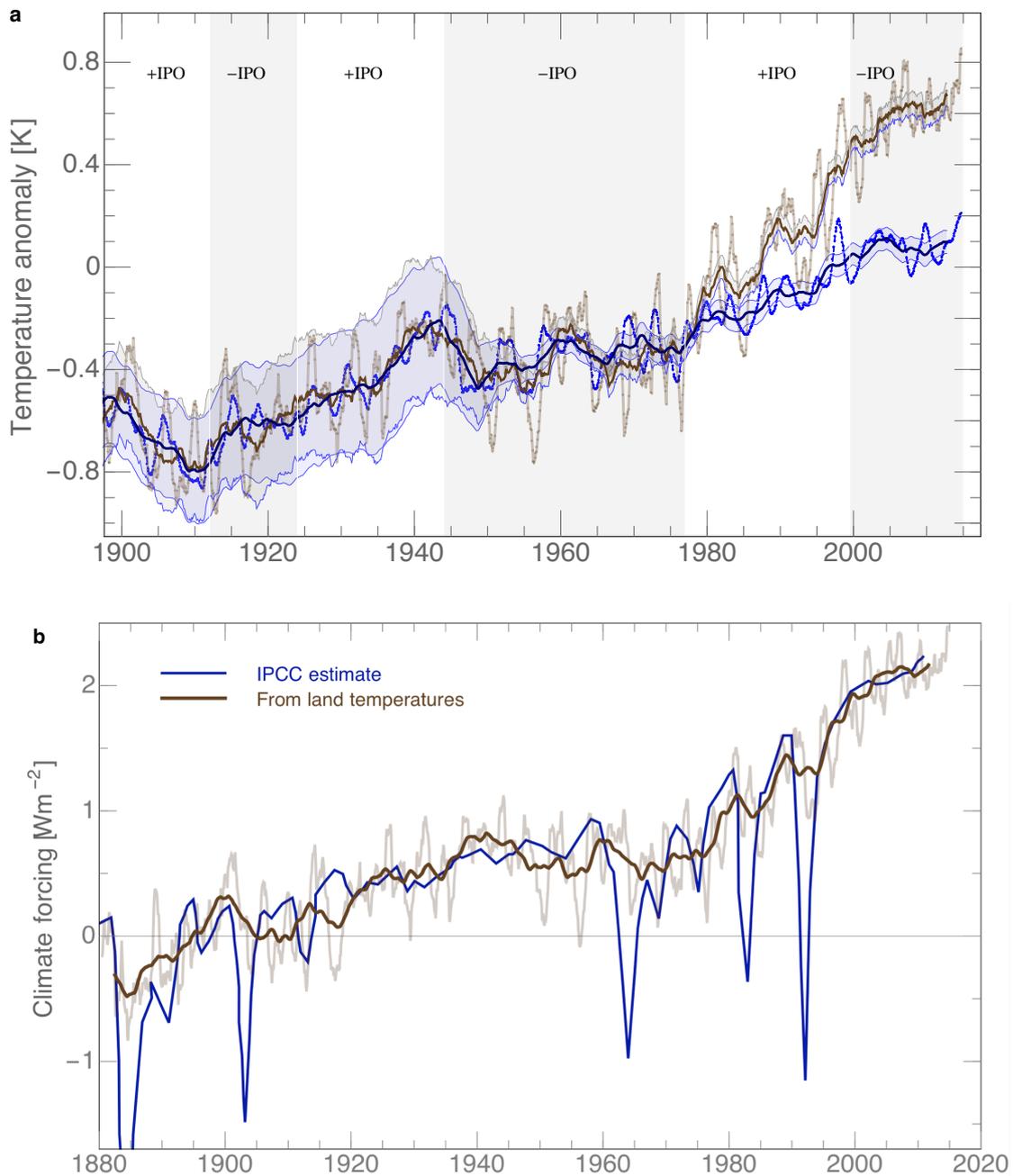

**Fig. 1.** Surface temperature anomalies and climate forcing. (a) Land (brown) and ocean (blue) ERSST v3 .5.4 temperature anomalies shown as one-year and 5-year running means. Ocean data are adjusted down by 0.16 degrees to match the land data in the reference period 1900-1970. Approximate IPO phases are marked following Henley et al. (2015). (b) Scaled land temperature anomaly from (a), with very minor changes accounting for ground absorption of heat, shown together with the IPCC consensus estimate of the total climate forcing (Myhre et al., 2013).

**2. Land surface temperature anomaly and climate forcing**

At any particular time, the surface warming anomaly $\Delta T(t)$ depends on the climate forcing energy flux $F(t)$ and the net inward energy flow $N(t)$. In the commonly used linear approximation (Murphy et al., 2009)

$$F = \lambda \Delta T + N + \varepsilon, \qquad (1)$$

where $\lambda \Delta T$ is the net change in outgoing radiation due to the change in surface temperature and $\varepsilon$ is the more rapidly varying noise from natural events or fluctuations. Different methods estimate (Murphy et al., 2009) the value of the constant $\lambda$ as $1.25 \pm 0.5$ Wm$^{-2}$K$^{-1}$.

We use the land temperature record (Fig. 1a) to fix the value of the global average climate sensitivity parameter $\lambda$. For land areas, the absorbed heat flux is almost two orders of magnitude smaller than the outgoing radiation flux (Beltrami, 2002; Hansen et al., 2011). We approximate the absorbed flux following the method of Beltrami (2002) and neglect the noise term. The resulting climate forcing flux is compared to the IPCC estimate (Myhre et al., 2013) in the Fig. 1b. In the present result, the strong minima due to volcano eruptions are not as prominent, but otherwise the two measures of the climate forcing are in good agreement. Both the climate forcing obtained from land surface temperatures and the consensus estimate (Myhre et al., 2013) indicate continuing increase in forcing over the recent 15-year period of slower warming. The value of $\lambda$ in the figure is 1.5 Wm$^{-2}$K$^{-1}$, well within the estimated range (Murphy et al., 2009). The same value is used throughout this work.

**3. SST anomaly and heat flux into the upper ocean**

At the global ocean surface, relating climate forcing to the average surface temperature change is more complicated because the radiated and absorbed energy fluxes are comparable in magnitude. Because of diffusive vertical heat transfer, the absorbed energy flux depends on the present and past SST. A linear approximation to this relationship can be evaluated using the ocean response function *G(t)*, which describes the variation of inward energy flux after a step change in the surface temperature at *t=0*. We describe the heat transfer via turbulent diffusive vertical mixing using the advection-diffusion equation (Munk, 1966)

$$\frac{\partial \theta(z,t)}{\partial t} = \frac{\partial}{\partial z}\left[\kappa(z)\frac{\partial \theta(z,t)}{\partial z}\right] - w\frac{\partial}{\partial z}\theta(z,t) . \qquad (2)$$

The temperature *Θ(z,t)* depends on depth and time, and the diffusivity *κ(z)* depends on depth *z*, while *w* is the constant upwelling velocity. In the steady state, the left-hand side is zero and the equilibrium temperature profile determines the ratio *κ(z)/w*. To determine the upwelling velocity *w* we use Argo seasonal temperature profiles and fit the variation in seasonal heat content calculated from equation (2) to the data. Once the functional form of *κ(z)* and the value of *w* are determined, we solve equation (2) to obtain the response function describing the heat flow into the ocean after a step change in the SST. The response function used here (Fig. 2a) is a global average, evaluated using only the measured Argo floats data without any free parameters. As the advection-diffusion equation is linear, from this function, we can calculate the response to an arbitrary sequence of changes in SST by adding responses to small annual or monthly changes. A much-simplified version of this approach was described in an earlier article (Marčelja, 2010), and details will be published separately.

As a test of the response function, the 0-700 m ocean heat content (OHC) was calculated from the SST shown in Fig. 1a and compared to the measured OHC (Fig. 2b). Over the most recent period, starting at around 1990 there is a very good agreement between different measurements and the calculation based on the SST and the model response function *G(t)*. Moreover, the results also agree with the OHC estimated from the changes of the sea level (Durack et al., 2014). Going further back in time the errors rapidly increase and different experimental estimates of OHC are no longer in agreement (Rhein et al., 2013). Our estimate of a larger OHC change in the period 1960-1990 is in line with the recent conclusion based on the consistency between sea level changes and the OHC and is supported by a range of model calculations. The source of the disagreement with earlier estimates was traced back to the sparse sampling of the Southern hemisphere heat content (Durack et al., 2014).

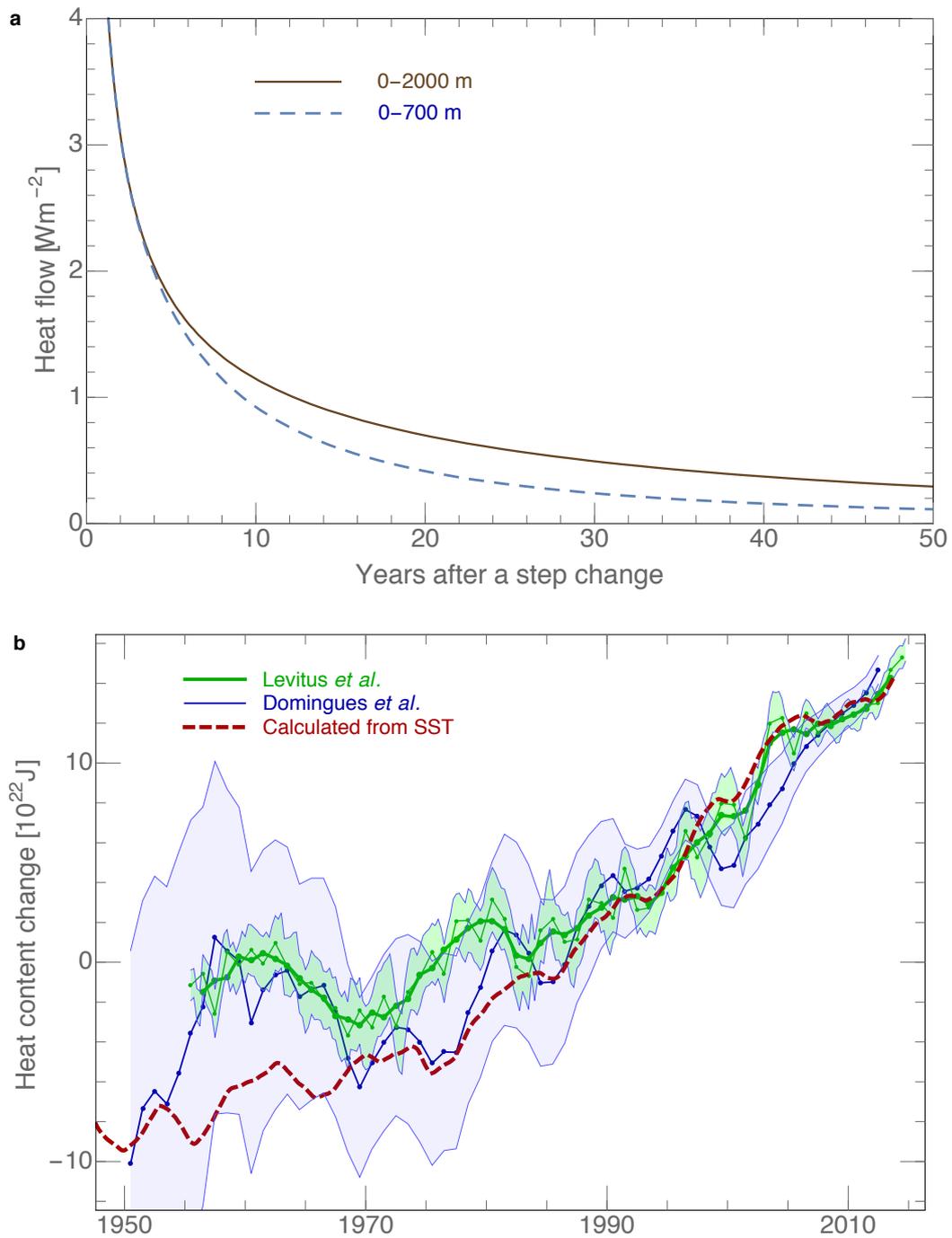

**Fig. 2.** Ocean response to surface heating. (a) The ocean response functions *G(t)* based on the turbulent diffusion approximation show the calculated heat flow into the top 700 m and top 2000 m after a one-degree step change of temperature at the surface. (b) The ocean heat content change 0-700 m evaluated with the SST values from Fig. 1 and the response function shown in (a) are compared with the data (Levitus et al., 2012, updated; Domingues et al., 2008, updated).

## 4. Evaluation of SST from the land surface temperature anomaly

In the last step, we use the climate forcing $F(t)$ deduced from the one-year running average of land surface temperatures (Fig. 1b) and the ocean inward heat flux response function $G(t)$ into the 0-2000 m depth range (Fig. 2a). These link the ocean and land surface temperatures via the linear approximation

$$F(t) = \lambda \Delta T_O + \int_{t'=0}^{t} G(t - t') \Delta T_O(t') \, dt'. \tag{3}$$

The ocean surface temperature anomaly $\Delta T_o$ was treated as an unknown and evaluated numerically from land surface temperatures and the upper ocean response function. In the discrete form, SST changes were calculated by progressively calculating shifts for each small step (in our case one month) in time. Monthly temperature changes were approximated as linear ramps. The solution depends on the initial value of climate forcing and the SST temperature anomaly prior to the start of the numerical integration. We started the integration in the year 1882, assuming the climate forcing evaluated from land temperatures as shown in Fig. 1b and the ocean temperature profile at equilibrium. The dependence on initial conditions progressively diminishes with passing years and by about 1930 it is negligible compared to SST uncertainties.

The comparison of the $\Delta T_o$ thus calculated with the measured SST (Fig. 1a) is shown in Fig. 3. If the model is consistent, the relations (1) and (3) hold; further if the values of $\lambda$ and the function $G(t)$ do not change with time the calculated and the measured ocean temperatures should match. The result shows good agreement between the calculated and measured ocean surface temperatures during the period 1930-2000. The modeling consistently relates the climate forcing to the land and ocean surface temperature changes over the period of $20^{th}$ century for which reliable data are available. The 1945-1975 pause in warming that coincided with a negative IPO phase is described correctly. The deviations during the strong El Nino years are not unexpected, because the model does not capture short-term fluctuations. However, from about the year 2000 the measured ocean surface warming slows down while the calculation based on the same inputs predicts continued strong warming. There were no events in the 1930-2000 period similar to the present slowdown in warming, which is seen as unique and unrelated to IPO oscillations.

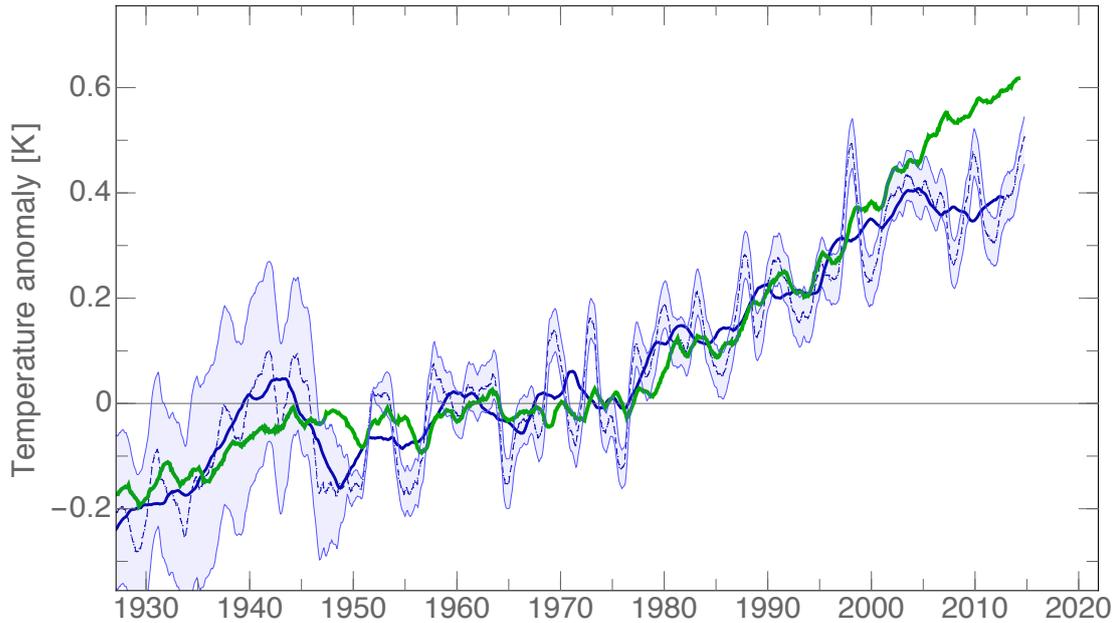

**Fig. 3.** The ocean surface temperature anomaly evaluated from land temperatures and the upper ocean response function. The SST calculated using the present model (green) and the ERSST data from Fig. 1a (blue) are shown as a one-year running mean and in the case of the data also as a 5-year running mean. The lack of agreement since the beginning of the slower warming period around the year 2000 indicates that the upper ocean response function is no longer sufficient to describe correctly the heat flow from the surface into the ocean interior.

Over the last 15 years for which the comparison fails, the climate forcing and correspondingly also the land surface temperatures behaved as expected on the basis of atmospheric composition (Myhre et al., 2013). Underestimated volcanic activity and reduced solar output (Schmidt et al., 2014) result in only a minor adjustment. In the same period, Argo data do not show any anomalies in the globally averaged heat content in the 0-700 m depth range. The ocean inward flux response function used in the modeling correctly describes the OHC in the same depth range (Fig. 2b). Significant changes in the ratio λ between the radiated energy and the SST change are not a plausible explanation. Variability linked to IPO cannot explain the result either, because the model held during the earlier cycles of IPO and failed only in the 21$^{st}$ century. These eliminations suggest that the recent slower warming is caused by a change in the ocean response to increasing SST. A change in deep ocean warming would not be captured by our turbulent diffusion response function, which was determined using the upper ocean data. Possible mechanisms contributing to the stronger heat subduction include intensifying westerly winds and vertical mixing in the Southern Ocean (Sheen et al., 2014) or increasing coastal upwelling that correlates with the increase in the land-sea temperature difference (Wang et al., 2015).

**5. Conclusion**

The failure of the present modeling since the beginning of this century is similar to the failure of all Coupled Climate Model Intercomparison Project ensemble means over the same period (e. g. Roberts et al., 2015). While model-based attribution of the global slowdown in warming to increased heat uptake by deeper ocean was proposed in earlier works (Meehl et al., 2011, Watanabe et al., 2013) these new findings are the result of a reliance on only the basic physical laws. The resulting transparency allowed us to understand the failure as a specific shortcoming of the models in describing the heat transfer into the deeper ocean. If future forecasting is to become more reliable, the deficiencies in modeling need to be understood and corrected, thus avoiding the addition of external constraints (Kosaka and Xie, 2013; Dai et al., 2015). If the increased deep ocean heat flow continues in line with the increase in climate forcing, the slower global warming might persist over the next decades. While the eventual equilibrium temperatures will not be affected, it should be prudent to anticipate the possibility of the continuation of the slower rate of global warming together with the seasonal aspects (Trenberth et al., 2014) of the changes in climate as experienced over the past decade.

**Acknowledgments**
ERSST Global monthly Land and Ocean Temperatures are available from NOOA at http://www1.ncdc.noaa.gov/pub/data/mlost/operational/products/. International Argo Program data are available at http://www.argo.ucsd.edu. Updated upper ocean heat content data are available at http://www.nodc.noaa.gov/OC5/3M_HEAT_CONTENT/basin_data.html (Levitus et al., 2012) and
http://www.cmar.csiro.au/sealevel/thermal_expansion_ocean_heat_timeseries.html
(Domingues et al., 2008).